\documentclass[12pt]{article}

\usepackage[ruled,vlined]{algorithm2e}
\usepackage{amsfonts}
\usepackage{amsmath}
\usepackage{amssymb}
\usepackage{amsthm}
\usepackage{array}
\usepackage{authblk}
\usepackage{bm}
\usepackage{booktabs}
\usepackage{caption}
\usepackage{color}
\usepackage{graphicx}
\usepackage{hyperref}
\usepackage{multirow}
\usepackage[authoryear,semicolon]{natbib}
\usepackage{physics}
\usepackage{siunitx}
\usepackage{standalone}
\usepackage{subfigure}
\usepackage{tikz}
\usepackage[normalem]{ulem}
\usepackage{url}

\newcolumntype{x}[1]{>{\centering\arraybackslash\hspace{0pt}}p{#1}}
\newcommand{\E}{\text{E}}

\newcommand{\Var}{\text{Var}}

\newcommand{\cov}{\text{cov}}

\newcommand{\betavec}{\boldsymbol{\beta}}
\newcommand{\epsilonvec}{\boldsymbol{\epsilon}}

\newcommand{\thetavec}{\boldsymbol{\theta}}

\newcommand{\Pivec}{\boldsymbol{\Pi}}
\newcommand{\Sigmavec}{\boldsymbol{\Sigma}}

\newcommand{\fvec}{\mathbf{f}}

\newcommand{\hvec}{\mathbf{h}}

\newcommand{\svec}{\mathbf{s}}
\newcommand{\uvec}{\mathbf{u}}
\newcommand{\vvec}{\mathbf{v}}

\newcommand{\xvec}{\mathbf{x}}

\newcommand{\Gau}{\text{Gau}}

\newcommand{\Avec}{\mathbf{A}}

\newcommand{\Dvec}{\mathbf{D}}
\newcommand{\Ivec}{\mathbf{I}}

\newcommand{\Qvec}{\mathbf{Q}}

\newcommand{\Xvec}{\mathbf{X}}
\newcommand{\Yvec}{\mathbf{Y}}
\newcommand{\Zvec}{\mathbf{Z}}

\theoremstyle{definition}

\usepackage{geometry}
\begin{document}

\def\spacingset#1{\renewcommand{\baselinestretch}%
{#1}\small\normalsize} \spacingset{1.5}


  \title{\textbf{\Large{Constructing Large Nonstationary Spatio-Temporal Covariance Models via Compositional Warpings}}}
  \author[1,*]{Quan Vu}
  \author[1,2]{Andrew Zammit-Mangion}
  \author[3]{Stephen J. Chuter}
  \affil[1]{\small{School of Mathematics and Applied Statistics, University of Wollongong, Australia}}
  \affil[2]{\small{Securing Antarctica’s Environmental Future, University of Wollongong, Australia}}
  \affil[3]{\small{Bristol Glaciology Centre, School of Geographical Sciences, University of Bristol, United Kingdom}}
  \affil[*]{\small{Corresponding author: quanv@uow.edu.au}}
  \date{}
  \maketitle

\begin{abstract}
Understanding and predicting environmental phenomena often requires the construction of spatio-temporal statistical models, which are typically Gaussian processes. A common assumption made on Gaussian processes is that of covariance stationarity, which is unrealistic in many geophysical applications. In this article, we introduce a deep-learning-inspired approach to construct descriptive nonstationary spatio-temporal models by modeling stationary processes on warped spatio-temporal domains. The warping functions we use are constructed using several simple injective warping units which, when combined through composition, can induce complex warpings. A stationary spatio-temporal covariance function on the warped domain induces covariance nonstationarity on the original domain. Sparse linear algebraic methods are used to reduce the computational complexity when fitting the model in a big data setting. We show that our proposed nonstationary spatio-temporal model can capture covariance nonstationarity in both space and time, and provide better probabilistic predictions than conventional stationary models in both simulation studies and on a real-world data set.
\end{abstract}


\spacingset{1.5} 

\section{Introduction}\label{sec:introduction}

There is often a need to build statistical models to explain and predict environmental phenomena. Statistical models have been used, for example, to obtain accurate estimates of Antarctic ice sheet mass balance and the resulting contribution to sea level rise \citep[e.g.,][]{Martin_2016}, and to determine sources and sinks of greenhouse gases \citep[e.g.,][]{Miller_2013}.
In addition to being representative of the underlying processes, statistical models used in the environmental sciences need to also be constructed in such a way that they can be fitted with big (e.g., remote sensing) data. In Section \ref{sec:application}, we provide an exemplar of such a case, where we predict changes in elevation in space and time from remote sensing data over Pine Island Glacier in Antarctica. This process is highly nonstationary since the glacier flows over complex bed topography with spatially-varying horizontal velocity.

Gaussian processes are widely used in spatial statistics to model environmental processes. However, Gaussian processes cannot be easliy used to model large data sets due to well-known issues concerning computational tractability when fitting and predicting \citep[see][for an example of recent work that attempts to circumvent computational difficulties in this context]{ExaGeoStat2018}. Therefore, a variety of alternative approaches have been developed  to model large spatial or spatio-temporal data. Some of the most commonly used approaches adopt low-rank representations; these include fixed rank kriging \citep{cressie2008fixed} and predictive processes \citep{banerjee2008gaussian}. Other approaches to model large spatial/spatio-temporal data use matrix sparsity to reduce the computational burden. These methods include sparse covariance methods such as covariance tapering \citep{kaufman2008covariance}, sparse precision methods based on stochastic partial differential equations \citep[SPDEs;][]{lindgren2011explicit}, and the class of so-called ``Vecchia approximations'' \citep{vecchia1988estimation}, which includes within it the nearest neighbor Gaussian process \citep{datta2016hierarchical} and the multiresolution approximation \citep{katzfuss2017multi, katzfuss2021general}. In general, low-rank approximations produce less accurate predictions than full-rank approximations do  in big-data settings \citep[e.g.,][]{heaton2019}, because they cannot capture the fine-scale behavior of the spatial fields.

Large spatial data offer the opportunity to model aspects of the process, such as nonstationarity, that are often difficult to characterize with small data sets. 
Nonstationarity occurs when the behavior of the process is not the same everywhere across the domain: In particular, covariance nonstationarity occurs when the covariance between two observations depend on the locations of the observations (and not just the displacement between the locations).
Some of the aforementioned approaches to model large spatial data do allow for nonstationarity \citep[e.g., fixed rank kriging models, ][]{cressie2008fixed}. However, the nonstationarity induced by these models tends to be too simplistic for  modeling real-world data.
Models that are regarded as reasonably flexible include low-rank models constructed using process convolutions with spatially varying kernels \citep{higdon2002space}, which have spatially varying model parameters, \citep[e.g.,][]{katzfuss2013bayesian}, and models built on domain partitions \citep[e.g.,][]{heaton2017nonstationary}.

One of the most popular approaches to modeling nonstationary \emph{spatial} data is the deformation approach, which was first proposed by \cite{sampson1992nonparametric}. The nonstationary behavior of spatial processes often arises from the properties of the geographical domain; the deformation approach acknowledges this by modeling nonstationarity through deformation of the spatial domain. There have been several works using deformations to model large nonstationary data. One of these is the work by \cite{hildeman2021deformed}, in which deformations are embedded in the SPDE model of \cite{lindgren2011explicit}. The deformation function in this model is not parameterized directly, which leads to difficulties in model interpretability. Another work is by \cite{guinness2021gaussian}, in which the warping function is a linear combination of the gradients of spherical harmonic functions. Usually, when constructing deformation models, the deformation functions need to be complicated enough to capture all the nonstationarity in the data, and the deformation functions should also be injective to avoid space-folding, which may otherwise lead to poor prediction performance. A recent work by \cite{azm2019deep} addressed these issues by using compositional injective warping functions to model large nonstationary spatial data. However, they used a low-rank structure on the warped domain which, as mentioned above, can lead to over-smooth predictions when one has big data \citep[e.g.,][]{heaton2019, zammit2017frk}. In this article, we use a similar approach to construct our warping function, but use a high-rank model (by using Vecchia approximations) on the warped domain.

Modeling nonstationary \emph{spatio-temporal} data is also of interest in several scientific domains. Spatio-temporal models are often categorized into two groups: descriptive models and dynamic models \citep{cressie2011statistics}, and nonstationarity has been considered in both cases. For example, descriptive nonstationary models have been proposed by \cite{garg2012learning} and \cite{salvana2020nonstationary}, both of which use convolutions to induce spatially-varying parameters. Examples of dynamic nonstationary models can be found in the work by \cite{huang2004modeling}, who use covariates in the covariance model, and in that by \cite{jurek2021multi}, who use multi-resolution decompositions of covariance matrices. 
The deformation approach to model nonstationary spatial data can be readily extended to model nonstationary spatio-temporal data. Some earlier works using the deformation approach in the spatio-temporal setting involve the dynamic model of \cite{morales2013state}, in which deformation was used to model spatial nonstationarity. \cite{shand2017modeling} extended the work by \cite{bornn2012modeling} and used dimension expansion (which can be cast as a deformation approach) to model spatial and temporal nonstationarity. \cite{salvana2020nonstationary} also used deformations to construct Lagrangian spatio-temporal covariances.

In this article, we extend the deep-learning-inspired deformation approach of \cite{azm2019deep} to a spatial-temporal context. Specifically, we construct descriptive nonstationary spatio-temporal models using deformation functions that warp the spatial domain and the temporal domain separately to address the nonstationarity in both space and time. Similarly to the spatial-only case, a stationary spatio-temporal covariance function is then used on the warped domain, and this covariance function can either be separable or nonseparable, resulting in a separable, nonstationary or a nonseparable, nonstationary covariance function on the original domain, respectively. 
A novel contribution in this article is the implementation of a sparse approximation for the precision matrix using the deep-learning package \textit{tensorflow} \citep{TensorFlowR} in \textit{R}, to reduce the computational complexity when estimating the spatial and temporal warping functions with big data. 
Our proposed models and implementation seek to address some of the limitations of previous works which, for example, either only considered nonstationarity in space \citep{hildeman2021deformed}, or only considered small (or subsampled) data sets \citep{shand2017modeling}.

The remainder of the article is organized as follows. In Section \ref{sec:model}, we introduce nonstationary spatial and spatio-temporal covariance models constructed through compositional warping functions. We also show how one can take advantage of the Vecchia approximation to make inference and prediction with the nonstationary models in a warped setting. In Section \ref{sec:sim_study}, we illustrate the implementation of the nonstationary spatio-temporal covariances on two simulated data sets: one with a separable covariance function, and one with an asymmetric covariance function. In Section \ref{sec:application}, we analyze the observed changes of ice surface elevation in Pine Island Glacier, Antarctica, for the period 2012--2017 using our proposed nonstationary covariance model. We show that the spatial warping function estimated by our model gives us information of spatial structure of the bed topography from measurements of elevation change over time, thus providing a physical interpretation to the inferred nonstationary behavior of the process. Code to reproduce results and figures in the manuscript can be found at \url{https://github.com/quanvu17/spatio_temporal_compositional_warpings}. Section 5 concludes.

\section{Model}\label{sec:model}

In this section, we first introduce the nonstationary spatial and spatio-temporal covariance models constructed via compositional warpings. We then show how parameter estimation and prediction with the warping model can be done via the Vecchia approximation. Finally, we briefly discuss aspects of model implementation.

\subsection{Nonstationary Spatial Covariances}

We begin by giving a brief overview of the deformation approach to modeling nonstationarity in the spatial case. Assume that we have noisy observations $\{Z_{k}: \ k = 1,\dots, n\}$ of a Gaussian process $\tilde Y(\cdot)$ indexed over a spatial domain $G_{\svec} \subset \mathbb{R}^2$ that we call the \textit{geographical spatial domain}. Specifically, let
\begin{equation}\label{eq:z_spatial_model}
	Z_{k} = \tilde Y(\svec_{k}) + \epsilon_{k};\quad k = 1,\dots,n,
\end{equation}
where $\svec_k \in G_{\svec}, k = 1,\dots,n$, $\{\epsilon_{k}: k = 1,\dots,n\}$ are Gaussian measurement errors which satisfy $\epsilon_{k} \sim \Gau(0, \tau^2)$, and $\tau^2$ is the variance of the measurement error.
We model the Gaussian process $\tilde Y(\cdot)$ as
\begin{equation}\label{eq:y_spatial_model}
	\tilde Y(\svec) = \xvec(\svec)' \betavec + Y(\svec), \quad \svec \in G_{\svec},
\end{equation}
where $\xvec(\cdot) = (x_1(\cdot), \dots, x_q(\cdot))'$ are $q$ covariates, $\betavec$ is a $q$-dimensional vector of unknown coefficients that need to be estimated, and $Y(\cdot)$ is a zero-mean Gaussian process.

An assumption often made with the Gaussian process $Y(\cdot)$ is that of (second-order) stationarity. A stationary Gaussian process has a covariance function that depends only on the displacement $\hvec = \svec - \uvec$ of the locations $\svec, \uvec$, that is,
\[ \cov(Y(\svec), Y(\uvec)) \equiv C_{G_{\svec}}(\svec, \uvec) = C^{o}_{G_{\svec}}(\hvec), \quad \svec, \uvec \in G_{\svec}, \]
where $C^{o}_{G_{\svec}}(\cdot)$ is a stationary covariance function on $G_{\svec}$.

In many environmental applications, it is often unrealistic to assume stationarity for the spatial process $Y(\cdot)$. We can use various methods to construct a nonstationary process. In this paper, we are particularly interested in using the deformation approach \citep{sampson1992nonparametric} since the inferred warping function can often be related to properties of the geophysical process, which in turn could be useful for model interpretability. 
In the deformation approach, one models the nonstationary covariance function on the geographical spatial domain $G_{\svec}$ as a stationary covariance function on a warped spatial domain $D_{\svec}$ through an injective and differentiable spatial warping function $\fvec_{\svec}: G_{\svec} \to D_{\svec}$. Then,
\begin{equation}\label{eq:cov_model_spatial}
	C_{G_{\svec}}(\svec, \uvec) = C_{D_{\svec}}(\fvec_{\svec}(\svec), \fvec_{\svec}(\uvec)) = C^{o}_{D_{\svec}}(\fvec_{\svec}(\svec) - \fvec_{\svec}(\uvec)), \quad \svec, \uvec \in G_{\svec},
\end{equation}
where $C^{o}_{D_{\svec}}(\cdot)$ is a stationary spatial covariance function on the warped spatial domain.

\subsection{Nonstationary Spatio-Temporal Covariances}\label{sec:nonstat_spattemp_cov}

We next extend the above spatial-only model to a spatio-temporal model. Now, the noisy observations $\{Z_{k}: k = 1,\dots, n \}$ are of a Gaussian process $\tilde Y(\cdot\,; \cdot)$ indexed over a geographic spatio-temporal domain $G_{\svec t} \equiv G_{\svec} \times G_t$, $G_t \subset \mathbb{R}$. Anologous to \eqref{eq:z_spatial_model}, we have that
\begin{equation}\label{eq:z_spatiotemp_model}
	Z_{k} = \tilde Y(\svec_{k}; t_k) + \epsilon_{k};\quad k = 1,\dots, n,
\end{equation}
and analogous to \eqref{eq:y_spatial_model}, we model the Gaussian process $\tilde Y(\cdot\,; \cdot)$ as
\begin{equation}\label{eq:y_spatiotemp_model}
	\tilde Y(\svec; t) = \xvec(\svec\,; t)' \betavec + Y(\svec; t), \quad \svec \in G_{\svec}, ~ t \in G_t,
\end{equation}
where $\xvec(\cdot; \cdot)$, $\betavec$ and $Y(\cdot; \cdot)$ are all defined similarly as in \eqref{eq:y_spatial_model}.

A stationary spatio-temporal Gaussian process, often assumed for its simplicity and computational tractability, has a covariance function that depends only on the spatial displacement $\hvec = \svec - \uvec$ of the spatial locations $\svec, \uvec \in G_{\svec}$, and the temporal displacement $w = t - v$ of the time points $t, v \in G_t$, that is, a covariance function that satisfies
\begin{equation} \label{eq:cov_spatiotemp_model}
	\cov(Y(\svec; t), Y(\uvec; v)) \equiv C_{G_{\svec t}}(\svec, \uvec; t, v) = C^{o}_{G_{\svec t}}(\hvec; w). 
\end{equation}

The deformation approach can be extended to model nonstationarity in both space and time.
Here, we propose to use different warping functions to warp the spatial domain and the temporal domain separately: a spatial warping function $\fvec_{\svec}: G_{\svec} \to D_{\svec}$, and a temporal warping function $f_t: G_{t} \to D_{t}$. Define the warped spatio-temporal domain as $D_{\svec t}$.
The covariance model then becomes
\begin{align}\label{eq:cov_model_spatiotemp}
	\begin{split}
		C_{G_{\svec t}}(\svec, \uvec; t, v) &= C_{D_{\svec t}}(\fvec_{\svec}(\svec), \fvec_{\svec}(\uvec); f_t(t), f_t(v)) \\ &= C^{o}_{D_{\svec t}}(\fvec_{\svec}(\svec) - \fvec_{\svec}(\uvec); f_t(t) - f_t(v)),
	\end{split}
\end{align}
for $\svec, \uvec \in G_{\svec}$; $t, v \in G_t,$ where $C^{o}_{D_{\svec t}}(\cdot; \cdot)$ is a stationary spatio-temporal covariance function on the warped domain.

To model complex spatial nonstationary behavior, we adapt the deep-learning-inspired deformation approach in \cite{azm2019deep}, where the spatial warping function is constructed from a composition of $L$ injective warping units (or layers), that is,
\begin{equation}\label{eq:warp_func_spatial}
	\fvec_{\svec}(\cdot) \equiv \fvec_L\,\circ\,\fvec_{L-1}\,\circ\,\cdots\,\circ\,\fvec_1(\cdot).   
\end{equation}
We consider axial warping units (which we describe below), and radial basis function units, which are described in \cite{azm2019deep}. This warping approach guarantees injection (which helps avoid space-folding and results in better predictive performance), and is relatively flexible when compared to other approaches often used in spatial applications, such as those based on splines.
The temporal domain is only one-dimensional, and therefore we can use an axial warping unit for the temporal warping function $f_t(\cdot)$, which is given by
\begin{equation}\label{eq:warp_func_temporal}
	f_t(\cdot) = \sum_{i = 1}^{r} w_i \phi_i(\cdot),
\end{equation}
where $w_i > 0, i = 1,\dots,r$, $\phi_1(t) = t; \phi_i(t) = \frac{1}{1 + \exp{-\theta_{1i}(t-\theta_{2i})}}, i = 2, \dots, r$, and $\theta_{1i}$ and $\theta_{2i}$ are parameters that are set such that $f_t(\cdot)$ can model reasonably complex functions on the entirety of $G_t$, as typically done when using basis functions to model spatial/temporal processes \citep[e.g.,][]{wikle2019spatio}. Note that the function $f_t(t), t \in G_t$, is strictly  increasing. This is to ensure that the order in time is preserved on the temporal warped domain: that is, to ensure that if $t_1 < t_2; t_1, t_2 \in G_t$ then $f_t(t_1) < f_t(t_2)$.
Note that it is possible for two (or more) warping functions to characterize the same covariance function. This identifiability problem is discussed elsewhere \citep[e.g.,][]{vu2021modeling}.

The spatio-temporal covariance model on the geographical domain can either be a separable or a nonseparable covariance function. A spatio-temporal covariance function is said to be separable if it can be written as the product of a spatial-only covariance function and a temporal-only covariance function, that is, if one can write down
\[ \cov(Y(\svec; t), Y(\uvec; v)) = C_{G_{\svec}}(\svec, \uvec) C_{G_t}(t, v), \quad \svec, \uvec \in G_{\svec};~ t, v \in G_t. \]
We note that,
\[ C_{G_{\svec}}(\svec, \uvec) C_{G_t}(t, v) = C_{D_{\svec}}(\fvec_{\svec}(\svec), \fvec_{\svec}(\uvec)) C_{D_t}(f_t(t), f_t(v)), \]
for $\svec, \uvec \in G_{\svec}$ and $t, v \in G_t,$ implying that, for the proposed covariance model in \eqref{eq:cov_model_spatiotemp}, the spatio-temporal covariance function on the geographical domain is separable if and only if the covariance function on the warped domain is separable.
If one anticipates having a nonseparable covariance function on the geographical spatio-temporal domain, then one needs to use a nonseparable covariance function on the warped domain. 

Spatio-temporal covariance functions are sometimes also assumed to be symmetric. In such cases, one can write
\[ \cov(Y(\svec; t), Y(\uvec; v)) = \cov(Y(\svec; v), Y(\uvec; t)), \quad \svec, \uvec \in G_{\svec}, ~ t,v, \in G_t. \]
When studying processes involving advection or directional flows, it is more appropriate to use asymmetric covariance functions. An asymmetric covariance function is always nonseparable \citep{gneiting2006geostatistical}. 
Therefore, if one anticipates having asymmetric covariances on the geographical spatio-temporal domain, then one needs to use an asymmetric covariance function on the warped domain. 

\subsection{Parameter Estimation and Prediction Using Vecchia Approximations}\label{sec:estimation}

Inference with the nonstationary deformation model can be done using restricted maximum likelihood (REML) as in  \cite{vu2021modeling}. Let $\Zvec \equiv (Z_{1},\dots, Z_{n})'$,
$\Yvec \equiv (Y(\svec_{1}; t_1),\dots, Y(\svec_{n}; t_n))'$,
$\Xvec \equiv (\xvec(\svec_{1}; t_1),\dots, \xvec(\svec_{n}; t_n))'$,
$\epsilonvec \equiv (\epsilon_{1},\dots, \epsilon_{n})'$,
and $\Sigmavec = \cov(\Yvec) + \cov(\epsilonvec)$. 
Let $\thetavec$ be the vector containing all parameters appearing in the covariance matrix $\Sigmavec$, which include those needed to construct the warping function and the covariance function on the warped domain.
To estimate the parameters, we maximize the log restricted likelihood for the model \eqref{eq:z_spatiotemp_model}--\eqref{eq:cov_spatiotemp_model}, which is given by
\begin{equation}\label{eq:reml}
	\mathcal{L}(\thetavec; \Zvec) = -\frac{n - q}{2} \log(2 \pi) + \frac{1}{2} \log \abs{\Xvec' \Xvec} + \frac{1}{2} \log \abs{\Qvec} - \frac{1}{2} \log \abs{\Xvec' \Qvec \Xvec} - \frac{1}{2} \Zvec' \Pivec \Zvec,
\end{equation}
where
\( \Pivec = \Qvec - \Qvec \Xvec (\Xvec' \Qvec \Xvec)^{-1} \Xvec' \Qvec \) and $\Qvec \equiv \Sigmavec^{-1}$ is the precision matrix of $\Zvec$.

Since one requires factorization of $\Qvec$ to evaluate \eqref{eq:reml}, the computational complexity of evaluating the log restricted likelihood is $O(n^3)$.
Since computations may become infeasible when $n$ is large, an approximation of $\Qvec$ is often needed. In this article, we use the Vecchia approximation \citep{vecchia1988estimation, katzfuss2021general}, to obtain a sparse approximation of $\Qvec$.

Consider the joint distribution of the observations, which can be written as the product of conditional distributions, that is,
\begin{equation}\label{eq:NNGP}
	p(\Zvec) = \prod_{i = 1}^{n} p(Z_i \mid \Zvec_{A(i)}) = \prod_{i = 1}^{n} \mathcal{N}(\mu_i, \Sigma_i),
\end{equation}
where $A(i) \equiv \{1, \dots, (i - 1)\}, i \ge 2$, is the set of all indices from 1 to $(i -1)$ and $A(1) = \varnothing$; 
$\mu_1 = \xvec(\svec_1, t_1)' \betavec$, $\Sigma_1 = \Sigma_{1, 1}$;
$\mu_i = \xvec(\svec_i, t_i)' \betavec + \Sigmavec_{i, A(i)} \Sigmavec_{A(i), A(i)}^{-1} (\Zvec_{A(i)} - \Xvec_{A(i)} \betavec)$, and $\Sigma_i = \Sigma_{i, i} - \Sigmavec_{i, A(i)} \Sigmavec_{A(i), A(i)}^{-1} \Sigmavec_{A(i), i}$. Here, we have used set subscripts to denote vector or matrix subsets. For example, $\Zvec'_{A(i)} \equiv (Z_1,\dots, Z_{i-1})'$, and similarly for matrices.

We obtain an approximation $\tilde p(\Zvec)$ of $p(\Zvec)$ by replacing the conditional distributions $\{p(Z_i \mid \Zvec_{A(i)})\}_{i = 2}^n$ in \eqref{eq:NNGP} with $\{p(Z_i \mid \Zvec_{N(i)})\}_{i = 2}^n$, where $N(i)$ is the collection of indices of the observations that are `nearest' (spatially/spatio-temporally) to $Z_i, i \ge 2$. Usually, one sets $\abs{N(i)}$ to $\textrm{min}(m,i - 1)$, where $m$ is the maximum number of neighbors considered for each datum \cite[e.g.,][]{datta2016hierarchical}. Then, instead of having to deal with a large covariance matrix for each conditional distribution, that is, the matrix $\Sigmavec_{A(i), A(i)}$, we only need to consider a smaller covariance matrix $\Sigmavec_{N(i), N(i)}$. The approximate joint distribution $\tilde p(\Zvec)$ has a sparse precision matrix $\tilde \Qvec$, where 
$\tilde \Qvec \equiv \tilde \Sigmavec^{-1} = (\Ivec - \Avec)' \Dvec^{-1} (\Ivec - \Avec),$ and
\begin{equation}\label{eq:A_matrix}
	\Avec = \left(
	\left\{
	\begin{aligned}
		&A_{1, j} = 0; \quad j = 1,\dots,n \\
		&\Avec_{i, N(i)} = \Sigmavec_{i, N(i)} \Sigmavec_{N(i), N(i)}^{-1}; \quad i = 2, \dots, n \\
		&A_{i, j} = 0; \quad j \notin N(i); \quad i = 2, \dots, n
	\end{aligned}
	\right.
	\right),
\end{equation}
\begin{equation}\label{eq:D_matrix}
	\Dvec = \left(
	\left\{
	\begin{aligned}
		&D_{1, 1} = \Sigma_{1, 1} \\
		&D_{i, i} = \Sigma_{i, i} - \Sigmavec_{i, N(i)} \Sigmavec_{N(i), N(i)}^{-1} \Sigmavec_{N(i), i} ; \quad i = 2, \dots, n \\
		&D_{i, j} = 0; \quad j \neq i ; \quad i = 1, \dots, n
	\end{aligned}
	\right.
	\right),
\end{equation}
are a sparse matrix and a diagonal matrix, respectively \citep{finley2019efficient}. We then replace $\Qvec$ in \eqref{eq:reml} with the (approximate) sparse precision matrix $\tilde \Qvec$. The computational complexity of the likelihood evaluation using the approximate sparse precision matrix is $O(n m^2)$.

Because the approximation $\tilde p(\Zvec)$ depends on the neighbor sets $N(i), i \ge 2$, and these sets also depend on the ordering of the observations $\Zvec$, how we order the observations and find the neighbor sets has an impact on the accuracy of the approximation. \cite{guinness2018permutation} investigated a few approaches on ordering the observations, and found that ordering approaches such as maximum-minimum distance ordering or random ordering might be better than the often used strategy of coordinated-based ordering. 
Following the conclusions of \cite{guinness2018permutation}, we therefore use maximum-minimum distance ordering in space and a simple random ordering in time (since the data sets we consider only span a few time points) to re-order the observations in Sections \ref{sec:sim_study} and \ref{sec:application}. 
Once we suitably re-order the observations, we define the neighbor sets of the ordered observations. In a spatio-temporal setting, we need the neighbor sets to include observations from multiple time points, to adequately capture the temporal dependency. 
To do this, we first rescale the temporal coordinates with a scaling factor chosen so that the neighbor sets often include at least some observations at adjacent time points. We then find the nearest neighbors using Euclidean distances in three-dimensional space (via the spatial coordinates and the rescaled temporal coordinates). 
Other approaches that could be used include a ``simple neighbor selection'' approach \citep{datta2016nonseparable}, which uses a Cartesian product of spatial nearest neighbors and temporal nearest neighbors.
Because the model \eqref{eq:cov_model_spatiotemp} involves warpings of the spatial and temporal domains, ordering and finding the neighborhood of the observations can also be performed on the warped coordinates. Our preliminary experiments indicate that first fitting the model and then using neighbors of ordered observations on the warped domain for re-fitting the model does not substantially change the parameter estimates.

After finding the parameter estimates, we can use these estimates to obtain the prediction $Y_{*}$ at a new spatial location and time point $(\svec_{*}, t_{*})'$,
\begin{align}
	\begin{split}
		&\E(Y_{*} \mid \Zvec) = \xvec'_{*} \betavec + \Sigmavec_{*, N_{*}} \Sigmavec_{N_{*}, N_{*}}^{-1} (\Zvec_{N_{*}} - \Xvec'_{N_{*}} \betavec), \\
		&\Var(Y_{*} \mid \Zvec) = \Sigma_{*, *} - \Sigmavec_{*, N_{*}} \Sigmavec_{N_{*}, N_{*}}^{-1} \Sigmavec_{N_{*}, *},
	\end{split}
\end{align}
where $N_{*}$ is the collection of indices of the observations nearest to $(\svec_{*}, t_{*})'$. The set subscripts are used to denote vector or matrix subsets, and the subscript $*$ denotes the quantity evaluated at the new location and time point $(\svec_{*}, t_{*})'$. The nearest neighbors used for prediction can be found either on the original domain $G_{\svec t}$ or on the warped domain $D_{\svec t}$. In Section \ref{sec:sim_study_sep}, we show the difference in the predictive performance when finding $N_{*}$ using neighbors on $G_{\svec t}$ and $D_{\svec t}$. We find that using the neighbors on $D_{\svec t}$ (rather than on $G_{\svec t}$) results in better predictive performance. This is to be expected, since stationarity is assumed on the warped domain, finding the neighbor sets on the warped domain will likely lead to a better approximation of the likelihood than doing so on the original domain.

\subsection{Implementation in \textit{TensorFlow} }

\begin{figure}[t!]
	\centering
	
	\includegraphics[width=1\textwidth]{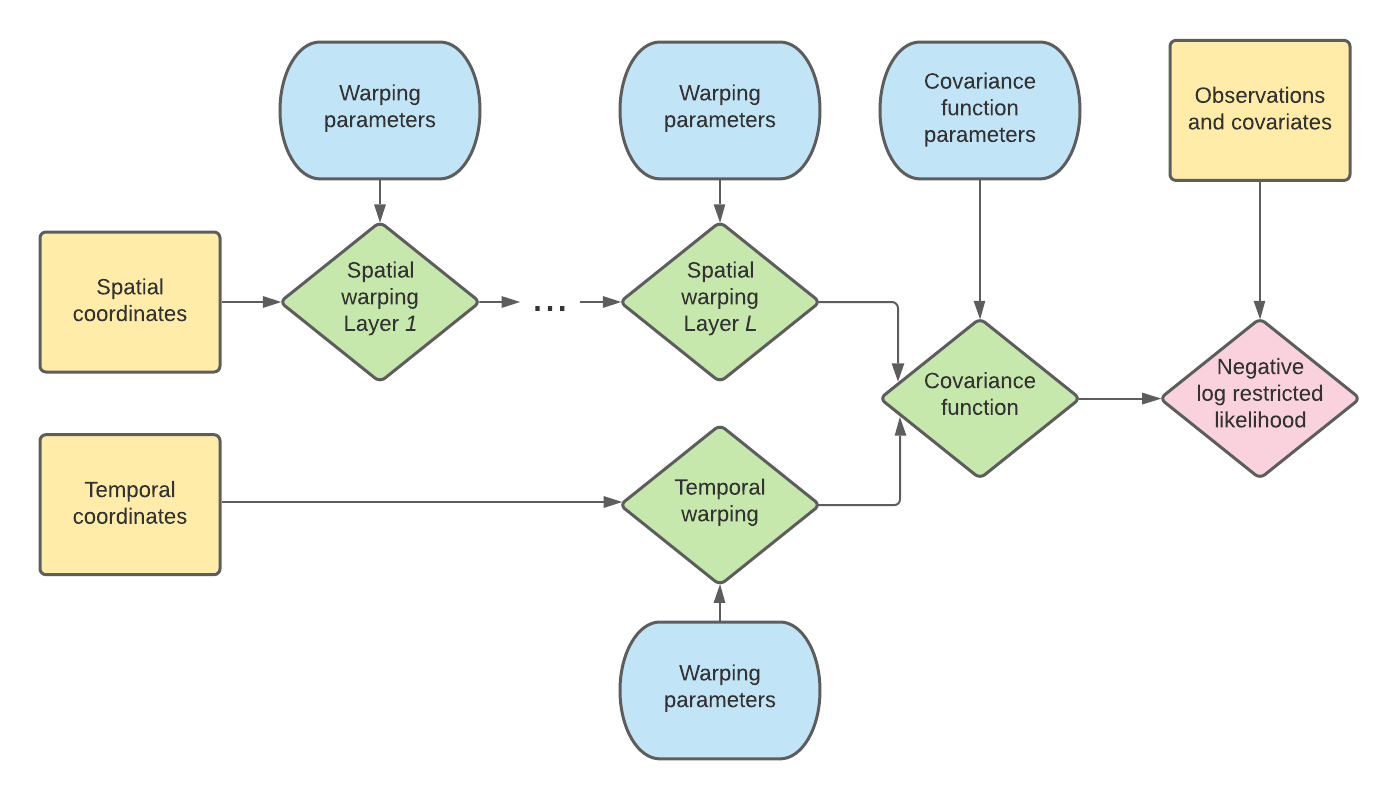}
	
	\caption{Model as constructed in \textit{TensorFlow}. Yellow: Inputs to the model. Blue: Parameters that need to be estimated.  Green: Functions constructed in \textit{TensorFlow}. Red: Loss function.}
	\label{fig:implementation}
\end{figure}

Since our model shares some similarities with traditional neural-network models, namely that both use compositions of functions to introduce model complexity, we used \textit{TensorFlow}, a deep-learning platform that is usually used for fitting neural-network models, in \textit{R} \citep{TensorFlowR} for model fitting.  We first constructed our model as a directed acyclic graph as shown in Figure \ref{fig:implementation}, where the spatial warping function has a multi-layer structure. The output of this graph is a loss function; in our case, it is the (negative) log restricted likelihood function. \textit{TensorFlow} minimizes the loss function by updating the parameters via a gradient-based optimizer. The gradients of the loss function with respect to each of the model parameters are computed through backpropagation \citep{rumelhart1986learning}, that is, the gradients are computed backwards from the loss function to the parameters on the graph using the chain rule. This approach allows one to implement and fit complicated compositional warping functions with relative ease.

To facilitate parallelization when model fitting, we developed new software for doing Vecchia approximations in \textit{TensorFlow}. Specifically, we first computed and stored the quantities required to construct $\Avec$ and $\Dvec$ in \eqref{eq:A_matrix} and \eqref{eq:D_matrix} as  three-dimensional tensors. These tensors were constructed such that the inner dimensions contain the required components (which are of size $1 \times 1$, $1 \times m$, or $m \times m$, as needed), and such that the outer dimension indexes the observation (and hence takes values in $i = 1, \dots, n$).  Matrix operations in \textit{TensorFlow}, such as matrix inversions and multiplications, using these three-dimensional tensors can be done in parallel in what is referred to as a batch matrix operation. This setup is particularly computationally advantageous when using a graphics processing unit (GPU). After the required matrix computations, the three-dimensional tensors are transformed into sparse two-dimensional tensors; these sparse tensors are then used to compute the log restricted likelihood in \eqref{eq:reml} using functions designed for sparse tensors in \textit{TensorFlow}.
Note that the \textit{TensorFlow} implementation of the Vecchia approximation can be easily adopted for other space-time warping approaches (e.g., one that warps space and time jointly).

All computations for the data examples in Sections \ref{sec:sim_study} and \ref{sec:application} were performed on a server with 64 cores in Intel Xeon E5-2683 @2.1 GHz processors, 256 GB RAM, and an NVIDIA GeForce GTX TITAN GPU. For a data set of 20,000 observations (such as in the simulation studies in Section \ref{sec:sim_study}), and choosing $m = 50$ neighbors, it takes around 10 minutes to fit the stationary model (that is, when the warping functions are fixed to identity functions), while it takes around 15 minutes to fit the nonstationary model.

\section{Simulation Studies}\label{sec:sim_study}

In this section, we demonstrate the use of our nonstationary spatio-temporal covariance model with two simulated data sets. The first data set was simulated using a separable covariance function, while the second one was simulated using an asymmetric covariance function.

\subsection{Simulation Study with Nonstationary Separable Covariances}\label{sec:sim_study_sep}

We begin with a simulation study with a data set from a nonstationary separable covariance function. The data were simulated from the Gaussian model \eqref{eq:z_spatiotemp_model} on an equally spaced $51 \times 51$ grid on the spatial domain $G_{\svec} = [-0.5, 0.5] \times [-0.5, 0.5]$, at 10 equally spaced time points on the temporal domain $G_t = [-0.5, 0.5]$, with zero mean and the covariance function \eqref{eq:cov_model_spatiotemp}. The spatial warping function, $\fvec_{\svec}(\cdot)$, was set to be a composition of two axial warping units (one for each spatial dimension), followed by a single resolution radial basis function unit, while the temporal warping function, $f_t(\cdot)$, was set to be an axial warping unit. The data were simulated from a stationary process, with the separable covariance function,
\begin{equation}\label{eq:sep_cov_model}
	C^{o}_{D}(\hvec; w) = \sigma^2 \exp(-a_{\svec} \norm{\hvec}) \exp(-a_t \abs{w}),
\end{equation}
at locations on the warped domain, which were then mapped backwards to the geographical domain to induce nonstationarity.

We randomly subsampled 80\% of the simulated data (i.e., 20,808 data points) and used these as training data for fitting the models. The remaining 20\% (i.e., 5,202 data  points) were used as validation data. For each validation data set, we first ordered the observations using maximum-minimum distance ordering in space \citep{guinness2018permutation}, and used $m = 50$ nearest neighbors in space and time to fit and predict with both the stationary and nonstationary covariance models. For the stationary covariance model we used \eqref{eq:cov_model_spatiotemp} with both $\fvec_{\svec}(\cdot)$ and $f_t(\cdot)$ set as identity functions. For the nonstationary covariance model we used \eqref{eq:cov_model_spatiotemp} with the spatial warping function $\fvec_{\svec}(\cdot)$ a composition of two single resolution radial basis function units (note that this is a misspecified warping function), and the temporal warping function $f_t(\cdot)$ an axial warping unit. For the covariance function on the warped domain we used \eqref{eq:sep_cov_model}. We predicted with the nonstationary model using two methods, where we (i) used the set $N_{*}$  of nearest neighbors on $G_{\svec t}$ or (ii)  used the set $N_{*}$  of nearest neighbors on $D_{\svec t}$. The process of simulating, fitting, and predicting the validation data was repeated 30 times. 

\begin{figure}[t!]
	\centering
	
	\includegraphics[width=0.8\textwidth]{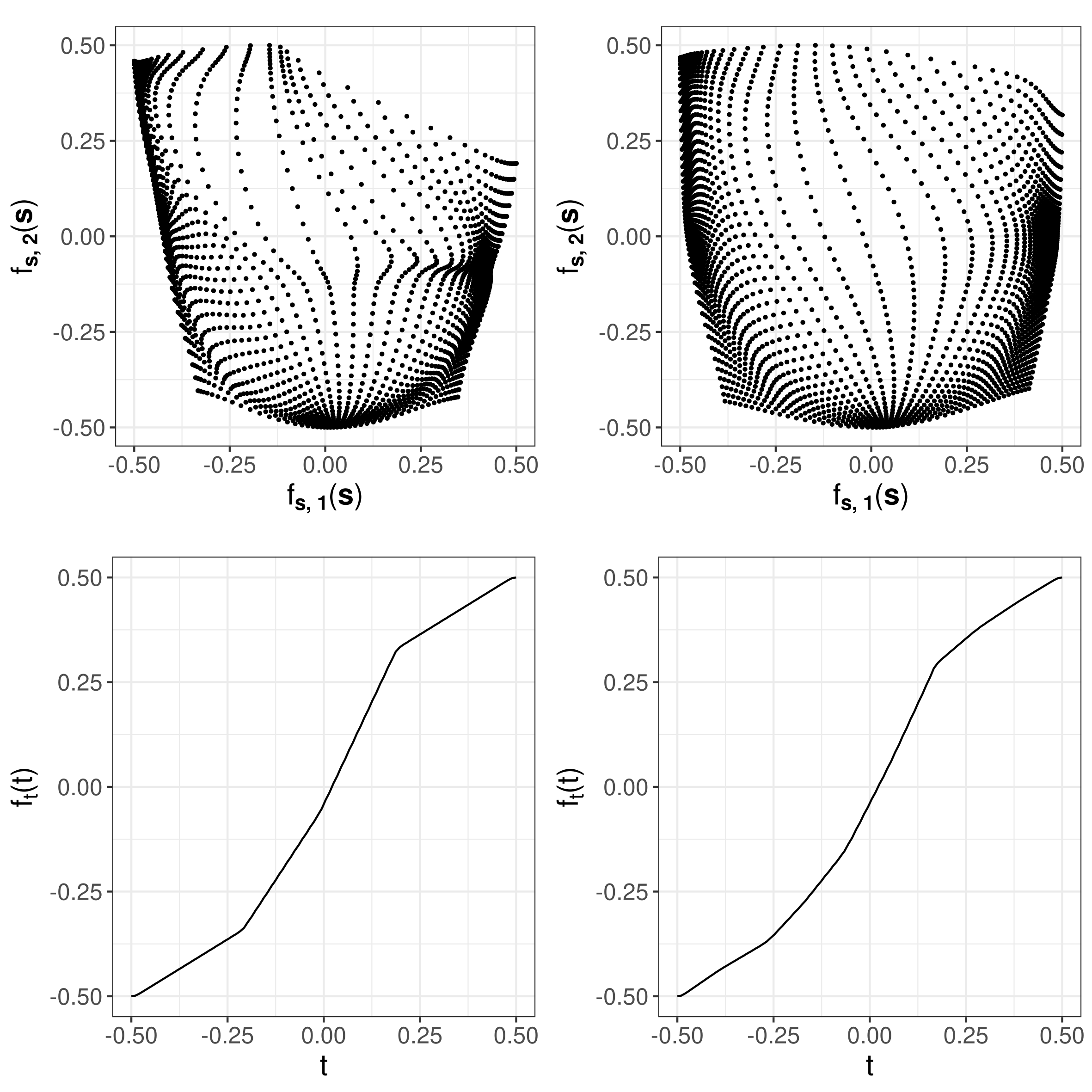}
	
	\caption{
		True and estimated spatial and temporal warping functions for the simulation study of Section \ref{sec:sim_study_sep}. Top row: True (left) and estimated (right) spatial warping function. Bottom row: True (left) and estimated (right) temporal warping function.
	}
	\label{fig:simstudy1_warpings}
\end{figure}

The predictive performance over the 30 sets of predictions was evaluated using two scoring rules: the root mean squared prediction error (RMSPE) and the continuous ranked probability score (CRPS) \citep{gneiting2007strictly}. Table \ref{tbl:sim1} shows the results from the simulation experiment. We can see that the nonstationary model produces predictions that are generally better than the stationary model. We also see that when predicting with the nonstationary model, using the warped locations on $D_{\svec t}$ to find the set $N_{*}$ provides better predictions. We also compared the estimated spatial and temporal warping functions to the true warping functions, and found that they are very similar (see Figure \ref{fig:simstudy1_warpings}). This demonstrates the ability of the model to recover the underlying warping function, even when Vecchia approximations are used to characterize the process on the warped domain.

\begin{table}
	\centering
	\caption{Hold-out validation results for the simulation study in Section \ref{sec:sim_study_sep}.}
	\label{tbl:sim1}
	\bgroup
	\def\arraystretch{1}
	\begin{tabular}{ |c|c|c| }
		\hline
		Model & RMSPE & CRPS \\
		\hline
		Stationary, separable & 0.394 &	0.215   \\
		\hline
		Nonstationary, separable (NNs on $G_{\svec t}$) & 0.330 & 0.174   \\
		\hline
		Nonstationary, separable (NNs on $D_{\svec t}$) & 0.303 & 0.162   \\
		\hline
	\end{tabular}
	\egroup
\end{table}

\subsection{Simulation Study with Nonstationary Asymmetric Covariances}\label{sec:sim_study_asym}

We now turn to a simulation study where we fit data generated using a nonstationary asymmetric covariance function. Similar to Section \ref{sec:sim_study_sep}, the data set was simulated from the Gaussian model \eqref{eq:z_spatiotemp_model} on an equally spaced $51 \times 51$ grid on the spatial domain $G_{\svec} = [-0.5, 0.5] \times [-0.5, 0.5]$, at 10 equally spaced time points on the temporal domain $G_t = [-0.5, 0.5]$, with zero mean and a covariance model \eqref{eq:cov_model_spatiotemp}. The spatial warping used, $\fvec_{\svec}(\cdot)$, was a single resolution radial basis function unit, while the temporal warping, $f_t(\cdot)$, was the identity function. The data were simulated from a stationary process, with the asymmetric, nonseparable covariance function,
\begin{equation}\label{eq:asym_cov_model}
	C^{o}_{D}(\hvec; w) = \sigma^2 \exp(-a \norm{\hvec - \vvec w}), 
\end{equation}
at locations on the warped domain, where $\vvec$ is a fixed velocity vector. 
These locations were then mapped backwards to the geographical domain to induce nonstationarity.
This covariance function has a mechanistic interpretation: it characterizes spatial fields moving in time according to $\vvec$. When combining the asymmetric covariance function with the warping function, the model can capture a process that is evolving according to a spatio-temporally varying vector field.

We used the simulated data to compare the predictive performance of the stationary and nonstationary covariance models, where the stationary model assumes that the warping functions are identity functions and where the nonstationary model has the same warping functions as in Section \ref{sec:sim_study_sep}. For both the stationary model and the nonstationary model we consider the covariance functions \eqref{eq:sep_cov_model} and \eqref{eq:asym_cov_model}. As in Section \ref{sec:sim_study_sep}, we randomly subsampled 80\% of the simulated data and used the remaining 20\% as validation data. Fitting and predicting was done using $m=50$ neighbors, and the process of simulating, fitting, and predicting the validating data was repeated 30 times. 

Table \ref{tbl:sim2} shows the results from the simulation study. We can see that the nonstationary model produces predictions that are generally better than the stationary model, and the predictions from the asymmetric models are better than those from the separable models. We also checked the parameter estimates, and found that the estimated warping functions are similar to the true warping functions (see Figure S1 in the supplementary material). As alluded to above, although the true velocity vector $\vvec$ is spatially invariant on $D_{\svec t}$, it induces a non-homogeneous velocity field on $G_{\svec t}$ as a result of the warping function. Thus, this nonstationary model is able to capture spatially varying temporal dynamics; in Figure S2 in the supplementary material, we show that the true underlying velocity field is well-estimated in this experiment.

\begin{table}
	\centering
	\caption{Hold-out validation results for the simulation study in Section \ref{sec:sim_study_asym}.}
	\label{tbl:sim2}
	\bgroup
	\def\arraystretch{1}
	\begin{tabular}{ |c|c|c| }
		\hline
		Model & RMSPE & CRPS \\
		\hline
		Stationary, separable & 0.409 &	0.232   \\
		\hline
		Stationary, asymmetric & 0.405 & 0.229   \\
		\hline
		Nonstationary, separable & 0.397 & 0.224   \\
		\hline
		Nonstationary, asymmetric & 0.383 & 0.216   \\
		\hline
	\end{tabular}
	\egroup
\end{table}

\section{Application to Changes in Antarctica Elevation}\label{sec:application}

\subsection{Background}\label{sec:application_background}

Global land ice, including the Greenland and Antarctic ice sheets, are significant components of the global sea level budget, contributing around 44\% of the global mean sea level rise over the satellite altimetry observation era from 1992 to present \citep{WCRPSLB2018}. 
Accurate estimates of ice sheet mass balance are critical to quantifying this contribution to sea level rise. Monitoring the long term evolution of the ice sheet provides insights into the driving processes causing mass changes, namely ice dynamic imbalance (variations in ice flow) or surface processes (variations in snowfall). Ascertaining the relative contribution of each of these processes is key to better ascertaining its sea level contribution over the coming decades.

Measurements of changes in ice sheet elevation ($\Delta h /\Delta t$) by multi-mission satellite altimetry provide the longest continuous record of ice sheet volume change, and hence mass change \citep{sorensen2018, schroder2019}. These are typically point-referenced observations, with their spatio-temporal resolution defined by the satellites' orbital characteristics. To estimate integrated volume or mass change, the ice sheet elevation change must be predicted over the entire ice sheet domain. One challenge of conventional satellite altimetry is sparse coverage in regions of steep topography at the ice sheet margins, where high velocity outlet glaciers are responsible for some of the largest $\Delta h/\Delta t$ rates \citep{hurkmans2012, hurkmans2014}.
The Synthetic Aperture Radar Interferometric (SARIn) mode of CryoSat-2 (launched in 2010) has greatly improved observational coverage in these marginal regions, although sparse coverage still remains over mountainous regions such as the Antarctic Peninsula.

We analyze an elevation change data set of Pine Island Glacier to illustrate the use of our nonstationary spatio-temporal covariance models. Pine Island Glacier has been the subject of extensive research by geoscientists because of the region's topography and its susceptibility to marine ice sheet instability: a positive feedback mechanism of increased ice mass losses.
The CryoSat-2 $\Delta h/\Delta t$ data used in this study are from \cite{BamberDawson2020}. For full processing details, see \cite{BamberDawson2020}. To derive $\Delta h/\Delta t$ from the original spatio-temporally irregular elevation ($h$) observations, a plane fit approach was applied across a regularly spaced polar stereographic grid \citep{mcmillan2014} at a 4 km spatial resolution. The plane fit approach solves simultaneously for $\Delta h/\Delta t$ whilst also accounting for the variation in topography within each grid cell. Due to the long 369-day repeat cycle of the satellite, each annual $\Delta h/\Delta t$ was computed using a three-year moving window of data weighted temporally using a tri-cube function. The resulting data product is a set of regularly spaced $\Delta h/\Delta t$ estimates at a 4 km interval on a grid of size 65 $\times$ 70 covering the glacier, at an annual temporal resolution over the period 2012--2017 (a total of 27,300 data points).
We then randomly subsampled 80\% of the data set (which is complete over the considered time period) to create a training data set (see Figure S3 in the supplementary material); the remainder was left for validation.

\subsection{Results}\label{sec:results}

The main aim of this section is to check whether our proposed nonstationary model can capture the nonstationary behavior in space and time of surface elevation change.
We set the spatial warping function in our nonstationary model, $\fvec_{\svec}(\cdot)$, to be a composition of two single resolution radial basis function units, while we set the temporal warping function, $f_t(\cdot)$ to be an axial warping unit. In this application, we are interested in the year-to-year fluctuations of $\Delta h/\Delta t$; these will be highly nonstationary in space and time, but likely unaffected by the nonseparable, asymmetric nature of glacial flow that is only evident over large time scales. We therefore let the covariance function on the warped domain be symmetric and separable:
\begin{equation*}
	C^{o}_{D}(\hvec; w) = \sigma^2 \exp(-a_{\svec} \norm{\hvec}) \exp(-a_t \abs{w}). 
\end{equation*}
We fitted the model and predicted the validation data using the methodology outlined in Section~\ref{sec:estimation} with $m = 50$.

\begin{figure}[t!]
	\centering
	
	\includegraphics[width=1\textwidth]{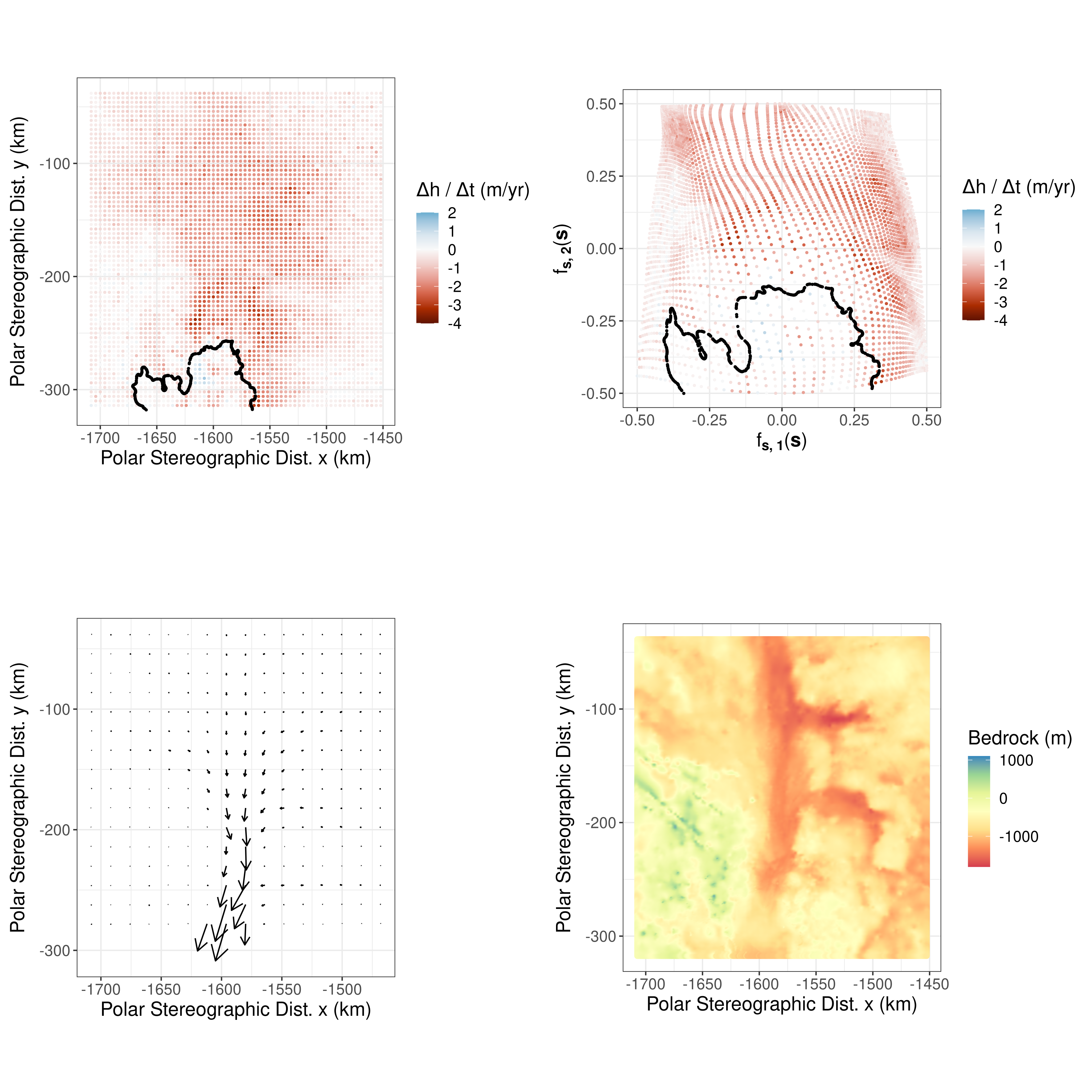}
	
	\caption{
		Estimated spatial warping function and geophysical properties of the spatial domain. Top left: Observation locations on the original domain. Top right: Locations on the warped domain. Bottom left: Ice velocity (the arrow size is proportional to the magnitude of the velocity; the largest arrow represents a velocity of approximately 4 km yr$^{-1}$). Bottom right: Bedrock topography in meters above sea level.
	}
	\label{fig:application_spatial_warpings}
\end{figure}

We first analyze the estimated spatial warping function in order to assess its characterization of the physical properties of the ice sheet domain. 
Figure \ref{fig:application_spatial_warpings} shows the estimated spatial warping function.
In general, a region on the geographical domain that has more variability will be expanded on the warped domain, while a region with lower variability will be contracted. The estimated warping function can therefore yield insights into the physical properties of different regions in our domain.
We see that there is a large expansion of the warped domain outside of the grounding line, which is defined as the line that separates the grounded ice from the ice shelf (floating ice), indicating that elevation change is more variable for floating ice than for grounded ice. This was expected:
Ice shelves differ from grounded ice in that they are freely floating on the ocean surface, therefore their flow characteristics are not governed by bedrock topography. Instead, the ice shelf provides buttressing back stresses to the grounded ice, regulating its flow into the ocean. When the ice is freely floating, elevation rates are much more spatially heterogeneous, in part due to the impact of basal melt caused by warmer ocean currents which has high spatial variability \citep{gourmelen2017channelized}.
We also see an expansion in the warped domain in the central part of the ice stream trunk. In this central part, the warping acts as a shear, in such a way that vertical lines in $G_{\svec}$ become diagonal in $D_{\svec}$ (see top right panel of Figure \ref{fig:application_spatial_warpings}). This warping is reflective of the ice velocity (bottom left panel of Figure \ref{fig:application_spatial_warpings}) which begins to change direction in this part of the domain. The high ice velocities \citep[which can reach 4 km per year;][]{Mouginot2014} can facilitate large changes in $\Delta h/\Delta t$. The sharp transition from compressed to expanded regions of the warped domain are also indicative of the steep bedrock topographic gradient of the ice stream (see bottom right panel in Figure \ref{fig:application_spatial_warpings}), which is grounded on much deeper bedrock than the rest of the drainage basin. It is this bedrock geometry that constrains the spatial structure of the grounded ice sheet flow. The combination of these two geophysical quantities, ice velocity and bedrock topography, allow for much greater spatio-temporal variations in ice elevation change in this region through changes in ice dynamics, modulated by the response of the ice shelf to changes in oceanic driven melt forcing. As such, the estimated warping function is encapsulating the large variations in geophysical processes occurring over short spatial length scales. 
In contrast, elevation change over the rest of the ice sheet is driven by small fluctuations in snowfall, leading to a more spatially homogeneous signal, and hence a contraction in the warped domain.
Note that information about ice velocity, bedrock topography and the grounding line is not used for estimating the warping function: the warping approach is able to capture physical properties of the region solely from the observations of elevation changes.

We next analyze the estimated temporal warping to assess the warping function's characterization of temporal nonstationarity.
We see that in the temporal dimension, the warped domain is contracted for the period 2012--2014, and expanded for the period 2014--2017 (see Figure S4 in the supplementary material).
This indicates that the temporal variability in ice sheet $\Delta h/\Delta t$ is modest for the earlier part of the time series, but larger after 2014. This corroborates with what is expected from geophysical considerations: For the 2010--2014 period, the elevation of the ice shelf was stable in comparison to the sustained thinning seen since the 1990s \citep{BamberDawson2020, Paolo2018}, implying a reduction in the oceanic driven basal melt in this region. The stability also coincides with minimal grounding line movement for this period \citep{Konrad2018, Joughin2016}. The result of this ice shelf stabilization is the regulation of the ice stream flow and a temporary state of disequilibrium, which results in smaller inter-annual variations in $\Delta h/\Delta t$. 
For the 2014--2017 period, the resumption of basal melt under the ice shelf at similar rates prior to 2010 \citep{BamberDawson2020, Paolo2018}, reduces the ice shelf's buttressing capability and therefore induces a resumption of accelerations in $\Delta h/\Delta t$ within the vicinity of the grounding line. Additionally, due to the material properties of ice, this rapid variation in flow regime will be propagated inland of the grounded ice sheet as a delayed response to the behavior at the grounding line. 

\begin{figure}[t!]
	\centering
	
	\includegraphics[width=1\textwidth]{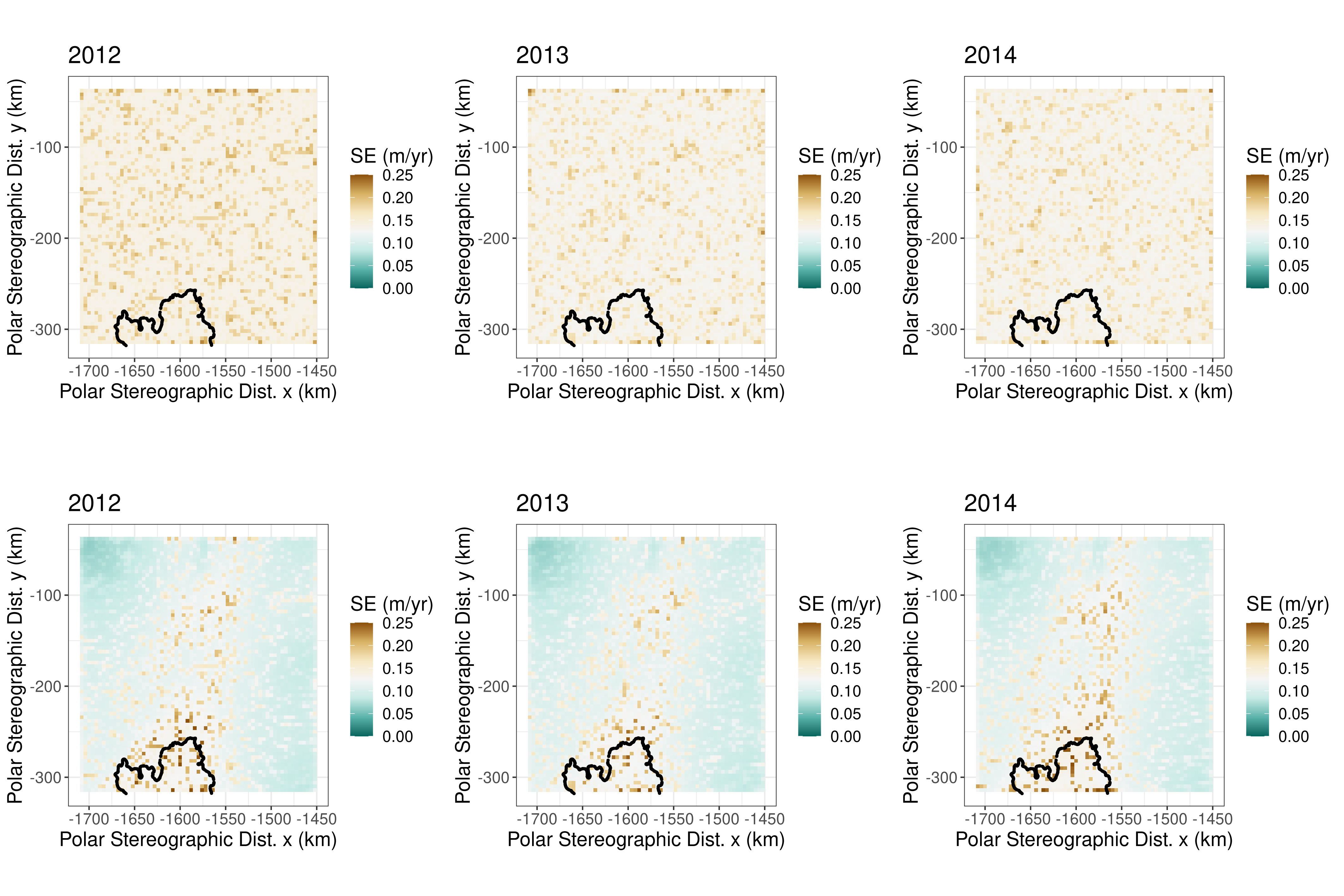}
	
	\caption{
		Prediction standard errors for the years 2012, 2013 and 2014 (left to right). Top row: Stationary model. Bottom row: Nonstationary model.
	}
	\label{fig:application_se}
\end{figure}

Finally, we compare the predictive performance of our nonstationary model with that of the stationary model in \eqref{eq:sep_cov_model} on the validation data. There was no noticeable difference between the predictions of the models, but the prediction standard errors from the two models differed substantially.
Figure \ref{fig:application_se} shows the prediction standard errors derived from the two models. We see that the prediction standard errors from  the stationary model have no clear pattern, since the observations are evenly distributed over the geographical domain.
Those from the nonstationary model, on the other hand, are related to the geography of the glacier: the prediction standard errors are higher outside of the grounding line and along the main outlet trunk of Pine Island Glacier, for the difference is that, for our nonstationary model, the prediction standard errors depend on the observed locations on the warped domain and not those on the geographical domain. In regions of contraction where there are several observations, the prediction standard errors are lower, while in regions of expansion (such as the region of the floating ice) where there are fewer observations, the prediction standard errors are larger.
We also notice that there is an improvement for the 95\% interval score \citep{gneiting2007strictly} from the nonstationary model (1.275) over that from the stationary model (1.362).
Therefore, in addition to the considerable interpretive advantages, the nonstationary model also yields slightly better uncertainty quantification than the stationary model.

\section{Conclusion}\label{sec:conclusion}

In this article, we introduce a new class of descriptive nonstationary spatio-temporal covariance function constructed through spatial and temporal warping functions. Specifically, separate spatial warping functions and temporal warping functions are used to capture covariance nonstationarity in both space and time. The warping functions are modeled as compositions of injective warping units, which allow for complex deformations. We then use a stationary spatio-temporal covariance function on the warped domain, which can be either separable or nonseparable, producing either a separable, nonstationary or a nonseparable, nonstationary covariance function on the original domain. Vecchia approximations are also used to reduce the computational complexity when fitting and predicting, when the data dimension is large.
We demonstrate the use of our proposed nonstationary spatio-temporal model on both simulated and real-world data sets. In general, we show that the spatial and temporal warpings can accurately characterize the nonstationary behavior of the processes, and can provide better probabilistic predictions than conventional stationary models.

There are several directions that can be considered for future work. 
First, in the article, we only consider warping space and time separately, and any nonseparability of the processes is captured by the covariance function. A different approach to model nonseparability is through a joint spatio-temporal warping function.
Moreover, the article considers a descriptive spatio-temporal model, which directly models the observations via spatio-temporal covariances. However, one my also consider dynamic spatio-temporal models, which treat the observations as a time series of spatial fields. Warping functions can then be used to model the nonstationarity in the spatial domain, while nonstationarity in the temporal domain can be addressed by other methods, such as time-varying parameters. 
Finally, we have demonstrated our approach with a real-world data set that is relatively spatio-temporally complete. We envision this model to be particularly useful for sparser data sets, such as historical (1992--2010) elevation change data sets, which precede the launch of the CryoSat-2 satellite.

\section*{Acknowledgements}

Q.V. was supported by a University Postgraduate Award from the University of Wollongong, Australia. A.Z.-M. was supported by the Australian Research Council (ARC) Discovery Early Career Research Award (DECRA) DE180100203 and by the ARC Special Research Initiative in Excellence in Antarctic Science (SRIEAS) Grant SR200100005 Securing Antarctica’s Environmental Future. S.J.C. was supported by the European Research Council (ERC) under the European Union's Horizon 2020 research and innovation programme under grant agreement No 694188 (GlobalMass) and by the European Space Agency (ESA) as part of the Climate Change Initiative (CCI) fellowship (ESA ESRIN/Contract No. 4000133466/20/I/NB).

\appendix

\setcounter{section}{0}
\setcounter{figure}{0}
\setcounter{table}{0}
\def\thesection{S\arabic{section}}
\def\thefigure{S\arabic{figure}}
\def\thetable{S\arabic{table}}

\section{Additional Figures}

\begin{figure}[h!]
    \centering
 	
 	\includegraphics[width=0.8\textwidth]{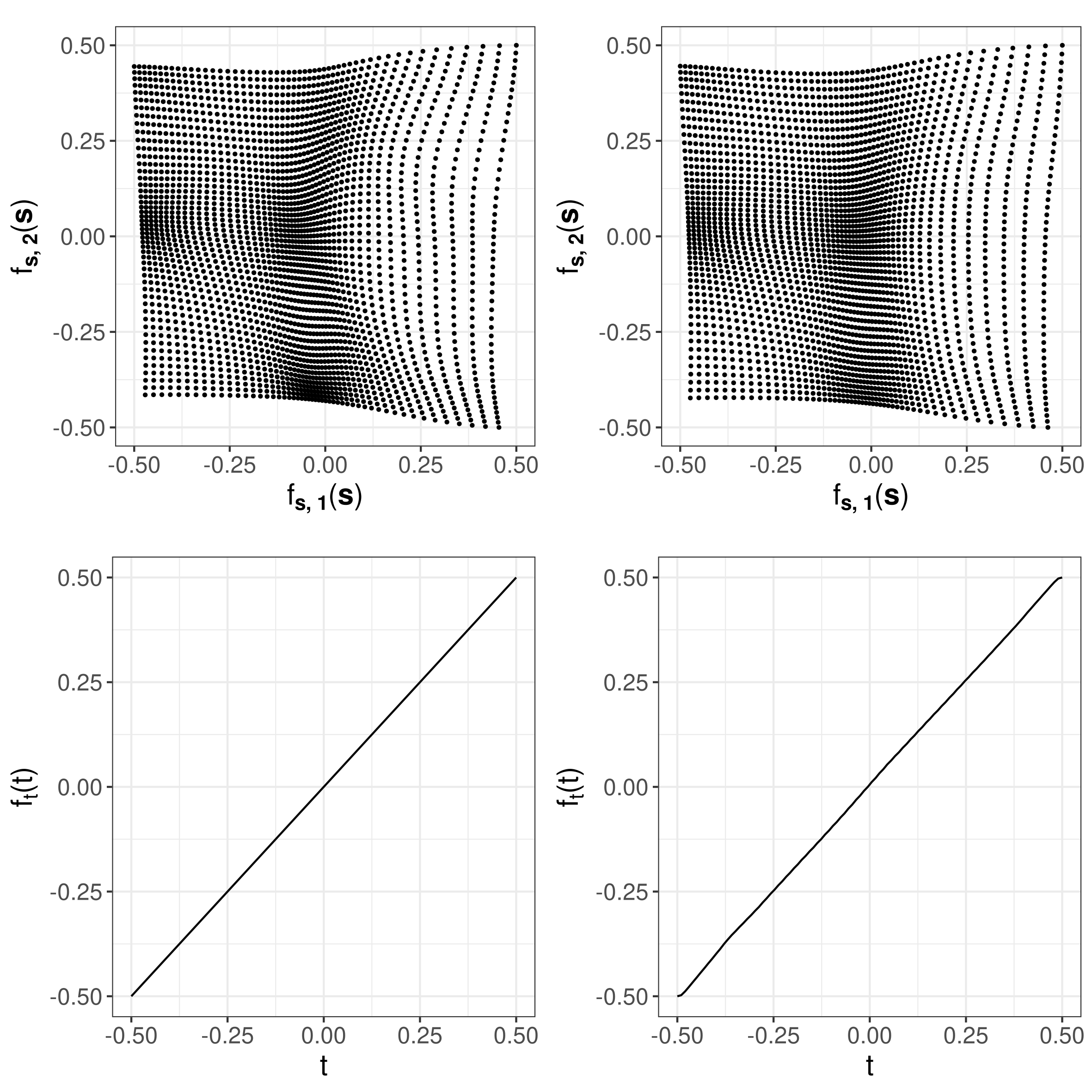}

 	\caption{
 	True and estimated spatial and temporal warping functions for the simulation study of Section 3.2. 
 	Top row: True (left) and estimated (right) spatial warping functions. Bottom row: True (left) and estimated (right) temporal warping functions.
 	}
 	\label{fig:simstudy2_warpings}
\end{figure}

\begin{figure}
    \centering
 	
 	\includegraphics[width=1\textwidth]{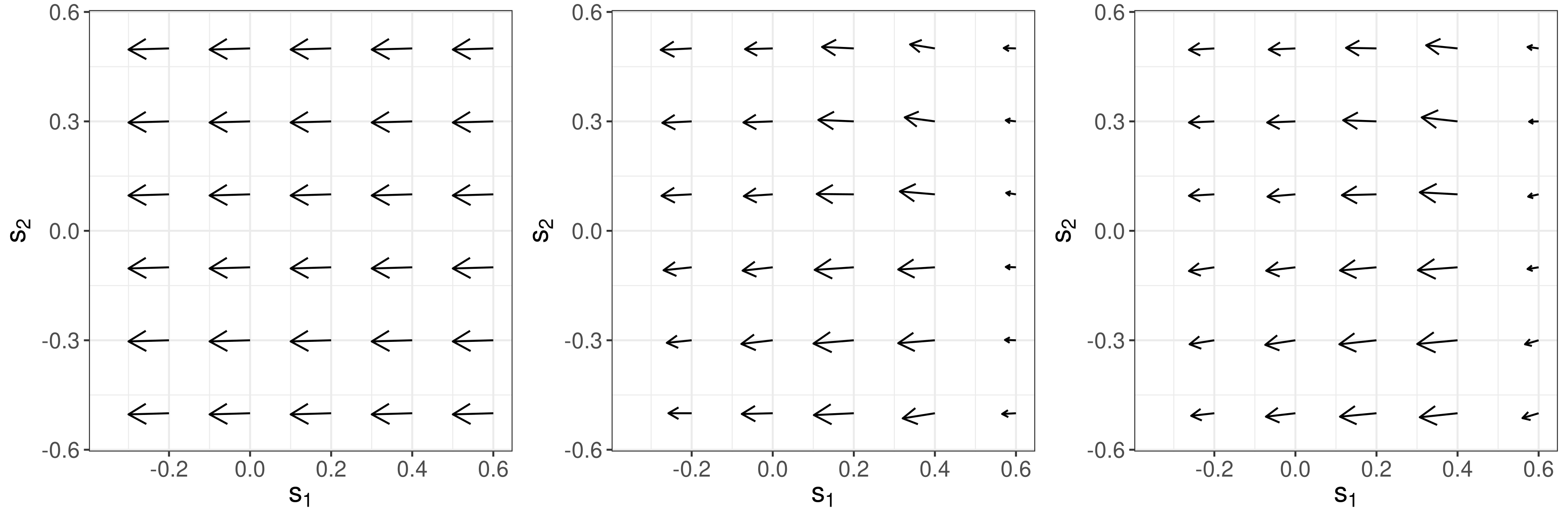}

 	\caption{
 	Velocity field evaluated at a few spatial locations on $G_{\svec t}$ for the simulation study of Section 3.2. 
 	Left and center panels: Velocity field as estimated by the stationary and nonstationary models, respectively. Right panel: The true underlying velocity field.
 	}
 	\label{fig:simstudy2_velocity}
\end{figure}

\begin{figure}
    \centering
 	
 	\includegraphics[width=1\textwidth]{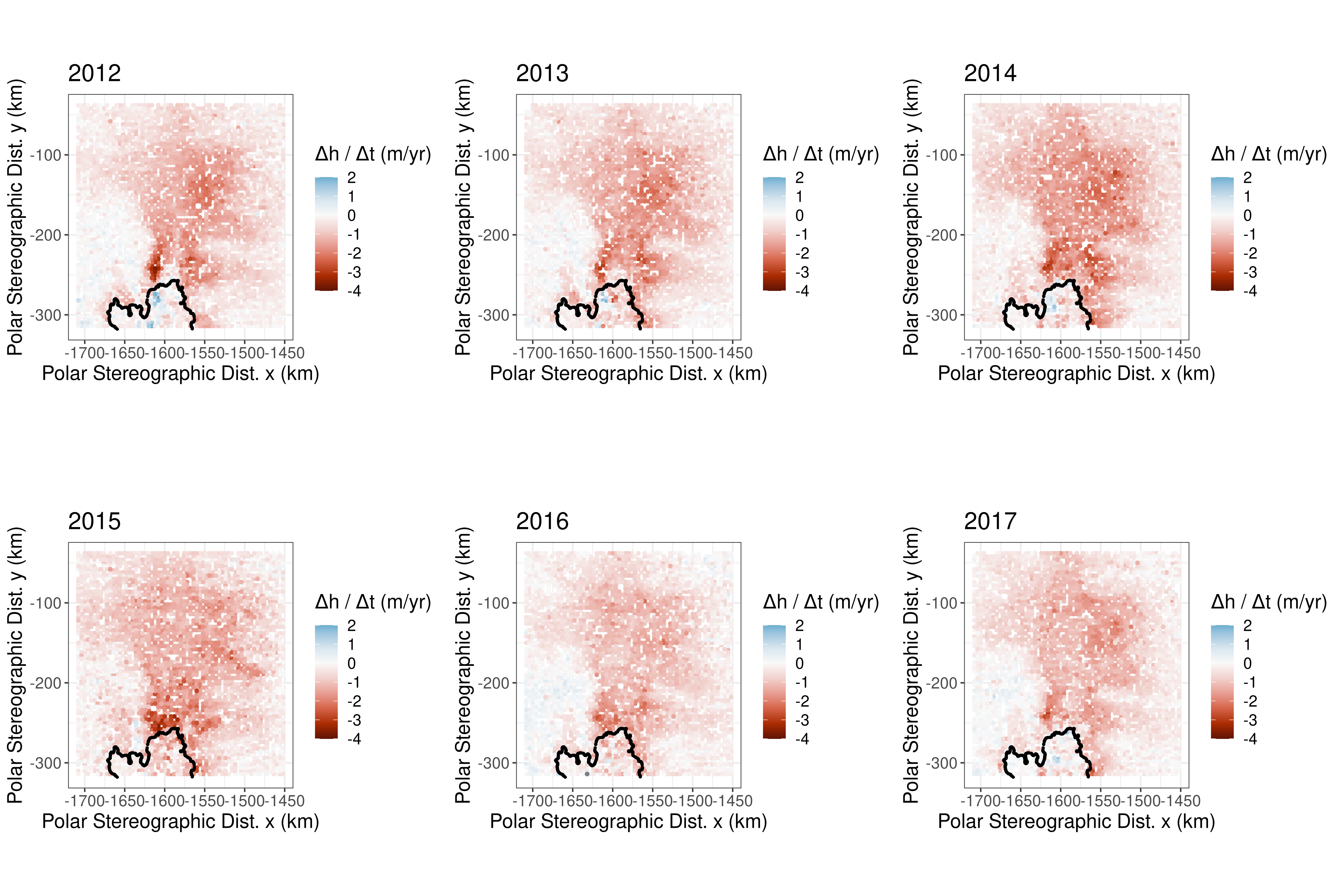}

 	\caption{
 	Observations of the elevation change in Pine Island Glacier in the period 2012--2017 that are used as training data.
 	}
 	\label{fig:application_obs}
\end{figure}

\begin{figure}
    \centering
 	
 	\includegraphics[width=0.4\textwidth]{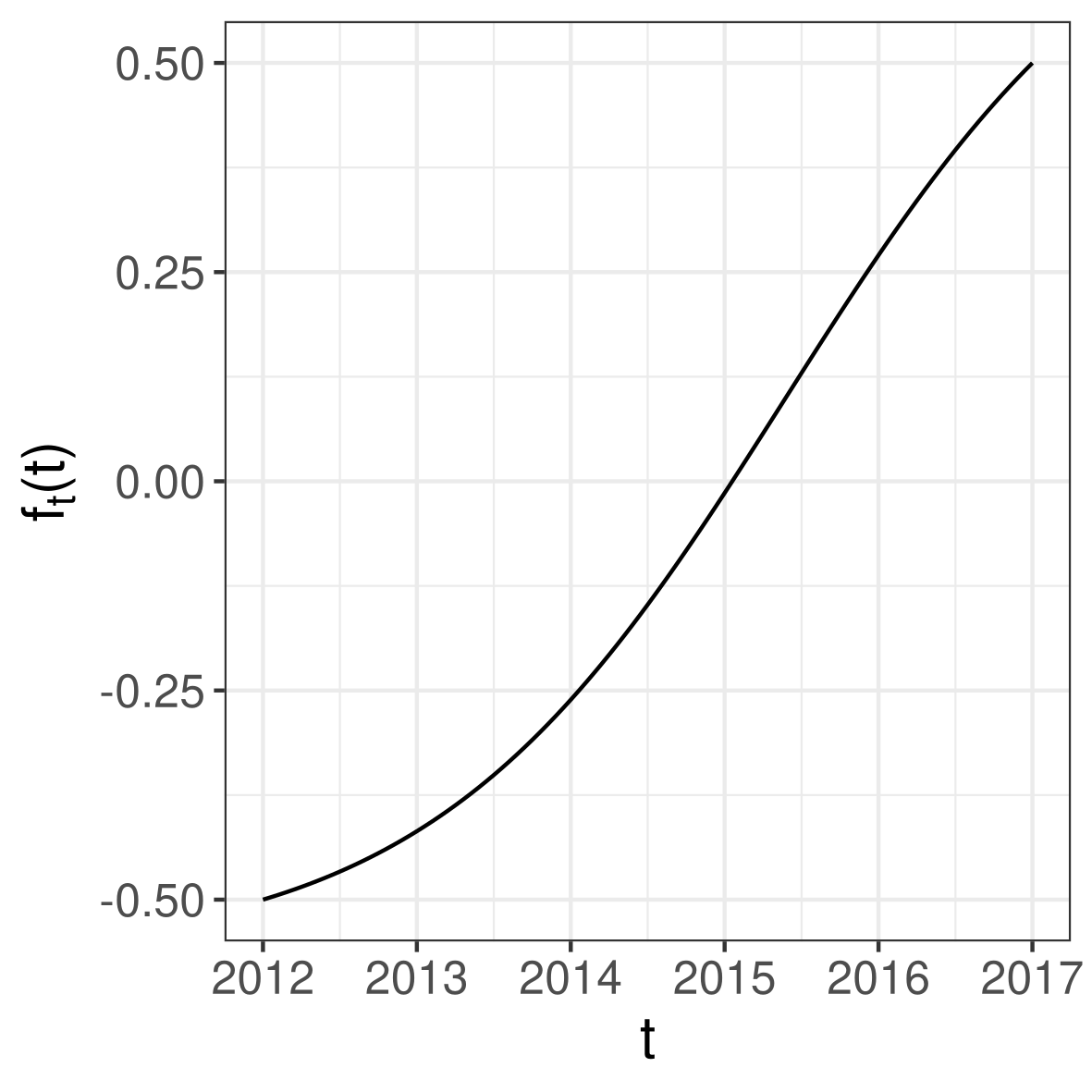}

 	\caption{
 	Estimated temporal warping function for the application case study of Section 4. 
 	}
 	\label{fig:application_temporal_warpings}
\end{figure}

\pagebreak
	
\bibliographystyle{apalike}
\bibliography{biblio}
	
\end{document}